\documentclass[reprint, superscriptaddress, secnumarabic, amssymb, nobibnotes, aps, prl]{revtex4-1}

\usepackage{graphicx}
\usepackage{epstopdf}
\usepackage[T1]{fontenc}
\usepackage[latin9]{inputenc}
\usepackage{amsbsy}
\usepackage{gensymb}
\usepackage[T1]{fontenc}
\usepackage[latin9]{inputenc}
\usepackage{amsmath}
\usepackage{amssymb}
\usepackage{bbm}
\usepackage{braket}
\usepackage{xcolor}
\allowdisplaybreaks
\usepackage{graphicx}
\usepackage[colorlinks=true]{hyperref}  
\hypersetup{
    bookmarks=true,         
    unicode=false,          
    pdftoolbar=true,        
    pdfmenubar=true,        
    pdffitwindow=false,     
    pdfstartview={FitH},    
    pdftitle={Type-II superconductivity in noncentrosymmetric superconductor PdTe$_{2}$},    
    pdfauthor={D. Singh, A. D. Hillier, Pabira K. Biswas,R. P. Singh},     
    pdfsubject={},   
    pdfcreator={},   
    pdfproducer={}, 
    pdfkeywords={} {} {}, 
    pdfnewwindow=true,      
    colorlinks=true,       
    linkcolor=blue, 
    citecolor=blue,        
    filecolor=magenta,      
    urlcolor=blue           
} 
\usepackage[normalem]{ulem}

\newcommand{\figref}[1]{Fig.~\ref{#1}}

\renewcommand{\approx}{\simeq}

\begin{document}
\title{\textrm{Coexistence of type-I and type-II superconductivity in topological superconductor PdTe$_{2}$}}
\author{D. Singh}
\affiliation{ISIS facility, STFC Rutherford Appleton Laboratory, Harwell Science and Innovation Campus, Oxfordshire, OX11 0QX, UK}
\author{Pabitra K. Biswas}
\affiliation{ISIS facility, STFC Rutherford Appleton Laboratory, Harwell Science and Innovation Campus, Oxfordshire, OX11 0QX, UK}
\author{Sungwon Yoon}
\affiliation{ISIS facility, STFC Rutherford Appleton Laboratory, Harwell Science and Innovation Campus, Oxfordshire, OX11 0QX, UK}
\author{C. H. Lee}
\affiliation{Department of Physics, Chung-Ang University, Seoul 06974, Republic of Korea}
\author{A. D. Hillier}
\affiliation{ISIS facility, STFC Rutherford Appleton Laboratory, Harwell Science and Innovation Campus, Oxfordshire, OX11 0QX, UK}
\author{R. P. Singh}
\affiliation{Indian Institute of Science Education and Research Bhopal, Bhopal, 462066, India}
\author{Amit}
\affiliation{Department of Physical Sciences, Indian Institute of Science Education and Research Mohali, Sector 81, S. A. S. Nagar, Manauli 140306, India}
\author{Y.Singh}
\affiliation{Department of Physical Sciences, Indian Institute of Science Education and Research Mohali, Sector 81, S. A. S. Nagar, Manauli 140306, India}
\author{K.-Y. Choi}
\affiliation{Department of Physics, Chung-Ang University, Seoul 06974, Republic of Korea}
\date{\today}
\begin{abstract}
\begin{flushleft}

\end{flushleft}
The type-II Dirac semimetal PdTe$_{2}$ was recently reported to be a type-I superconductor with a superconducting transition temperature $T_{c}$ $\approx$ 1.65 K. However, the recent results from tunneling and point contact spectroscopy suggested the unusual state of a mixture of type-I and type-II superconductivity. These contradictory results mean that there is no clear picture of the superconducting phase diagram and warrants a detailed investigation of the superconducting phase. We report here the muon spin rotation and relaxation ($\mu$SR) measurements on the superconducting state of the topological Dirac semimetal PdTe$_{2}$. From $\mu$SR measurements, we find that PdTe$_{2}$ exhibits mixed type-I/type-II superconductivity. Using these results a phase diagram has been determined. In contrast to previous results where local type-II superconductivity persists up to $H_{c2}$ $\approx$ 600 G, we observed that bulk superconductivity is destroyed above 225 G. 
\end{abstract}

\maketitle

\section{Introduction}
Topological superconductivity (TSC) is an exotic quantum state of matter that has been one of the most extensively studied research field in condensed matter physics owing to their novel electronic states \cite{AP,XL1,XL2}. The search for materials exhibiting topological superconductivity is fascinating, mainly motivated by the realization of the spin-triplet component of the superconducting order parameter, and the emergence of gapless surface quasiparticle states consisting of massless Majorana fermions \cite{MS,ZSM}. These Majorana fermions \cite{NR,AK} are of significant interest because of their potential applications in fault-tolerant quantum computation \cite{XL2,LFC,MZH}.\\ Recently, extensive efforts have been made to realize topologically non-trivial bands that might coexist with superconductivity \cite{ZL}.  Especially from a materials perspective, tremendous efforts have been made to fabricate such phase of matter through extrinsic doping and intercalation in the bulk topological insulators \cite{MPS1,MPS2,YSH}. Illustrative examples include Sr$_{x}$Bi$_{2}$Se$_{3}$ \cite{ZL} with $T_{c}$ $\sim$ 3.0 K, Cu$_{x}$Bi$_{2}$Se$_{3}$ \cite{YSH} with $T_{c}$ $\sim$ 3.5 K, Nb$_{x}$Bi$_{2}$Se$_{3}$ \cite{YQ} with $T_{c}$ $\sim$ 3.4 K, and Tl$_{x}$Bi$_{2}$Se$_{3}$ \cite{ZWANG} with $T_{c}$ $\sim$ 2.28 K. Various other routes were also pursued to stabilize the TSC in different material classes by taking advantage of the non-trivial topological nature of the electronic bands. For example, it has been argued that TSC can be induced by designing proximity-coupled heterostructures in topological surface states by interfacing with conventional superconductors \cite{CB,VS,PZ}, and by applying pressure on topological systems \cite{JLZ}. Likewise, superconductivity emerges in a confined dimension by forming a mesoscopic point contact between the pure element and topological materials \cite{LAA,HWH,SDL}. All these methods work towards a common goal to realize an ideal system in which topology and superconductivity can be studied unambiguously.\\
Lately, transition metal dichalcogenides attracted tremendous interest due to their rich structural chemistry and possible realization of topological non-trivial electronic band structures \cite{AA,HH,MY,MSB}. A well known example is PdTe$_{2}$, which is reported to be a type-II Dirac semimetal with a tilted Dirac cone \cite{MSB,YL,FF,HN,OJ}, exhibiting superconductivity below the critical temperature $T_{c}$ $\approx$ 1.6 K \cite{JG,FJ,HLC}. Notably, the superconducting phase in this compound emerges naturally without the aforementioned external tuning effects, serving as a promising candidate for topological superconductivity  \cite{FF}. Instigated by the topological nature of PdTe$_{2}$, Leng \textit{et al.} studied the detailed superconducting properties of PdTe$_{2}$ using the magnetic and transport measurements \cite{HLC}. Interestingly, experiments reveal type-I superconductivity in PdTe$_{2}$ with the thermodynamic field, $H_{c}$(0), yielded to be between 250 Oe to 3000 Oe, respectively \cite{HLC}. The observation of enhanced superconductivity in PdTe$_{2}$ was later attributed to the surface superconductivity. However, the experimentlly estimated critical field deviates from the standard Saint-James-de gennes surface sheath critical field \cite{HLC,DS,DS1}, suggesting that the observed behavior is not the typical case of surface superconductivity. This behavior was later linked to the topological nature of PdTe$_{2}$. Though the specific heat \cite{AY} and magnetic penetration depth \cite{MVS,STN} measurements confirm a conventional fully gapped s-wave superconductivity. Scanning tunneling microscopy (STM) and spectroscopy (STS) experiments further support the conventional superconducting gap in PdTe$_{2}$, which seems to rule out the possibility of topological superconductivity at the surface of PdTe$_{2}$\cite{SD}. Remarkably, the STM and STS experiments also reveal a mixed type-I and type-II superconductivity in PdTe$_{2}$, which opens up another possible case scenario for the observed behavior \cite{OJ,ASS}. In these studies, the superconducting phase diagram is not fully elucidated; however, they do indicate the formation of vortices at certain locations. Recently, soft point contact spectroscopy (PCS) data were also taken as evidence for mixed type-I and type-I superconductivity on the surface. The origin of mixed type-I/type-II superconducting behavior in PdTe$_{2}$ is attributed to intrinsic electronic inhomogeneity, where certain points on the surface of crystal fall in the type-II regime \cite{ASS,TLL}. The existence of both types of superconductivity in the same material is rare and of great experimental and theoretical interest. More importantly, it warrants a thorough investigation of the superconducting phase of PdTe$_{2}$ through microscopic tools like muon spin rotation and relaxation.\\ 
As muons are local probe of magnetism, this technique is well suited to determine the dominant internal field components in the superconducting state of the sample. In the past, this method has been used successfully to distinguish between type-I and type-II superconducting materials such as Sn (IV) \cite{VSE} and the recently published BeAu \cite{BA1,BA2}. By analyzing the $\mu$SR spectrum, one can probe different magnetic regions in the system. In particular, the ability to extract the magnetic field distribution from the $\mu$SR spectra allows to map regions of a Meissner state where the applied magnetic field is entirely expelled from the sample, an intermediate state where normal and Meissner states coexist, and a mixed state where the well-defined flux-line lattice (FLL) is formed.\\
In this work, we have utilized the transverse field (TF) muon spin rotation measurements to successfully construct the full superconducting phase diagram of PdTe$_{2}$. Our TF-$\mu$SR spectra at an applied field $H_{app}$ = 100 G demonstrates the internal field peaks due to both intermediate state and mixed state suggesting the coexistence of both type-I and type-II superconductivity. The superconductivity is completely suppressed at $H_{app}$ = 225 Oe, consistent with the earlier reports of bulk superconductivity state in PdTe$_{2}$ judged from susceptibility measurements \cite{HLC}. Measuring the collapse of the superconducting gap in STM measurements also reveals an upper critical field, $H_{c2}^{\perp}$ $\approx$ 200 Oe \cite{OJ}. This work is definitive in demonstrating that PdTe$_{2}$ is a mixed superconductor which lies on the boundary between type-I and type-II behavior and can be driven easily between the two states by external parameters, such as temperature and magnetic field. The admixed superconductivity in PdTe$_{2}$ is speculated to be due to the inhomogeneous distribution of local coherence length, $\xi$, originating from randomly distributed impurity/defects.
\begin{figure}
\includegraphics[width=1.0\columnwidth]{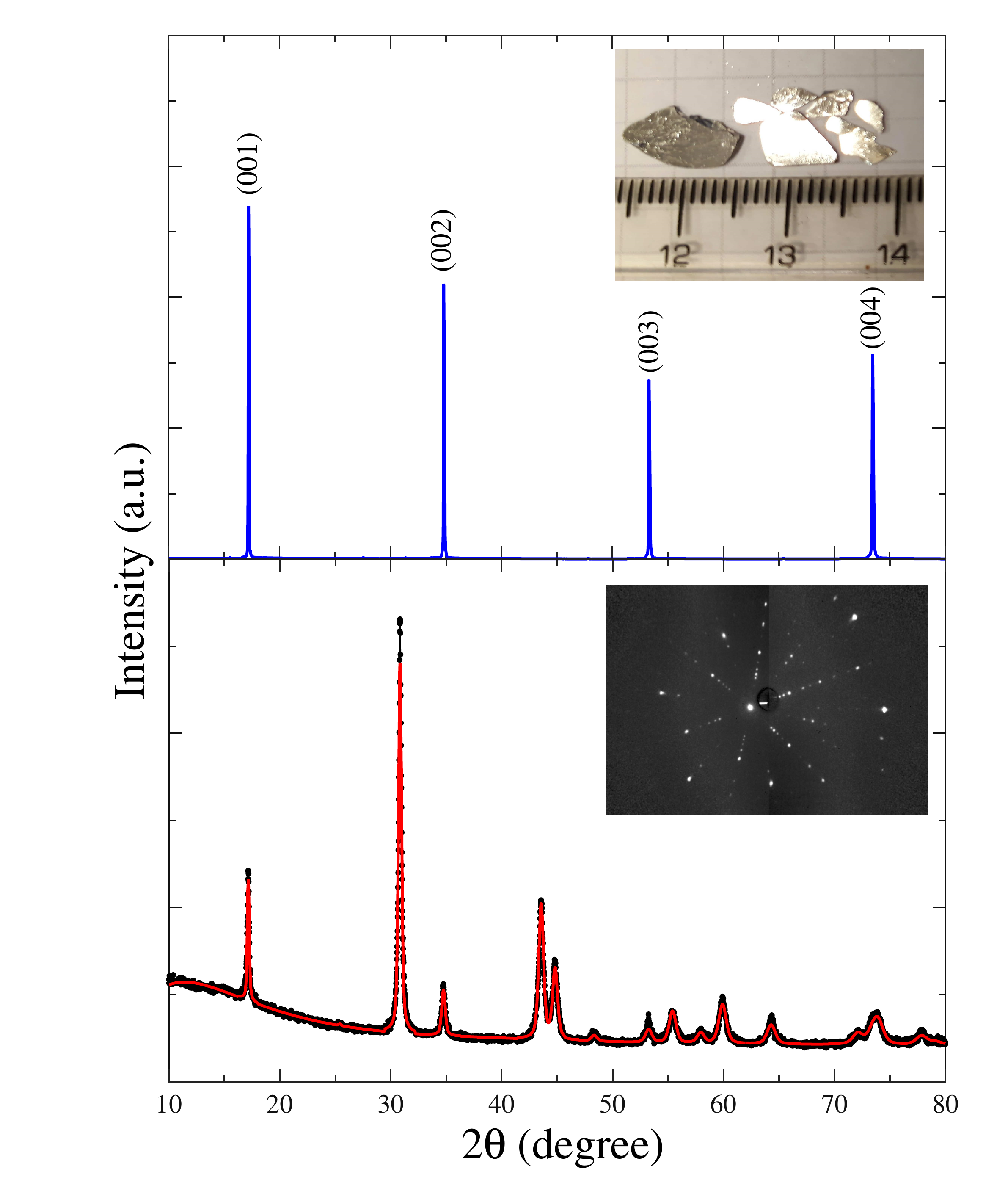}
\caption{\label{Fig1:Fig1}The single crystal and powder x-ray diffraction pattern of PdTe$_{2}$ at room temperature. The upper inset shows the optical image of PdTe$_{2}$ single crystal pieces. The bottom inset shows the Laue diffraction pattern of the PdTe$_{2}$ single crystal at room temperature.}
\end{figure}
\section{Results and Discussion}
High-quality single crystals of PdTe$_{2}$ were grown by the melt growth method as described in Ref. \cite{SD}. Figure \ref{Fig1:Fig1}(a) shows the x-ray diffraction (XRD) data along the basal plane reflections from PdTe$_{2}$. The sharp peaks along (00$l$) confirm the single crystalline growth of the sample. Laue backscattering further confirms the single-crystalline nature of the sample, as shown in the inset of Fig.\ref{Fig1:Fig1}(b). A few crystals [see inset Fig.\ref{Fig1:Fig1}(a)] were crushed into a very fine powder for XRD measurements which verify the phase purity and hexagonal CdI$_{2}$ structure (space group P$\bar{3}$$m$1) (no. 164) of PdTe$_{2}$ \cite{LT}. For $\mu$SR measurements, several single crystal pieces of PdTe$_{2}$ were mounted on a silver sample holder and placed in a dilution refrigerator operating in the temperature range of 0.05 K - 3 K. The $\mu$SR measurements were performed using the MuSR spectrometer at the ISIS pulsed muon facility, STFC Rutherford Appleton Laboratory, Didcot, United Kingdom \cite{SLL}. In the TF mode, an external magnetic field was applied perpendicular to the muon-spin direction. The magnetic field was applied in the range of 25 Oe $\le$ H$_{app}$ $\le$ 225 Oe above the superconducting transition temperature, $T_{c}$, of the sample and then cooled it to the base temperature. Muon spin rotates with the applied magnetic field and depolarizes as a consequence of magnetic field distribution inside the sample. The TF-$\mu$SR data were analyzed using the MaxEnt technique to determine the probability distribution, P(B), of the internal magnetic fields \cite{BDR}. In the zero field (ZF) muon spin relaxation measurements, the stray fields at the sample position due to neighboring instruments and the Earth's magnetic field are canceled to within $\sim$ 1.0 $\mu$T using three sets of orthogonal coils and an active compensation system. A full description of the $\mu$SR technique may be found in \cite{MSM}. 

ZF-$\mu$SR can detect internal magnetic fields as small as 0.1 G without applying external magnetic fields, making it a highly valuable tool for probing spontaneous magnetic fields due to TRS breaking in exotic superconductors.
 Hence, we performed the ZF-$\mu$SR measurements to search for the possible TRS breaking in the superconducting state of PdTe$_{2}$. This technique was proved to be very crucial in establishing the TRS breaking in several superconductors such as Sr$_{2}$RuO$_{4}$ \cite{GML,JXY}, LaNiC$_{2}$ \cite{ADH}, and Re$_{6}$X(X = Ti,Hf,Zr) \cite{RPS,DSJ,DSS}. Figure \ref{Fig1:Fig1} shows the ZF-$\mu$SR spectra for PdTe$_{2}$ at (a) 0.1 K, and (b) 2.5 K respectively.
\begin{figure}
\includegraphics[width=1.0\columnwidth]{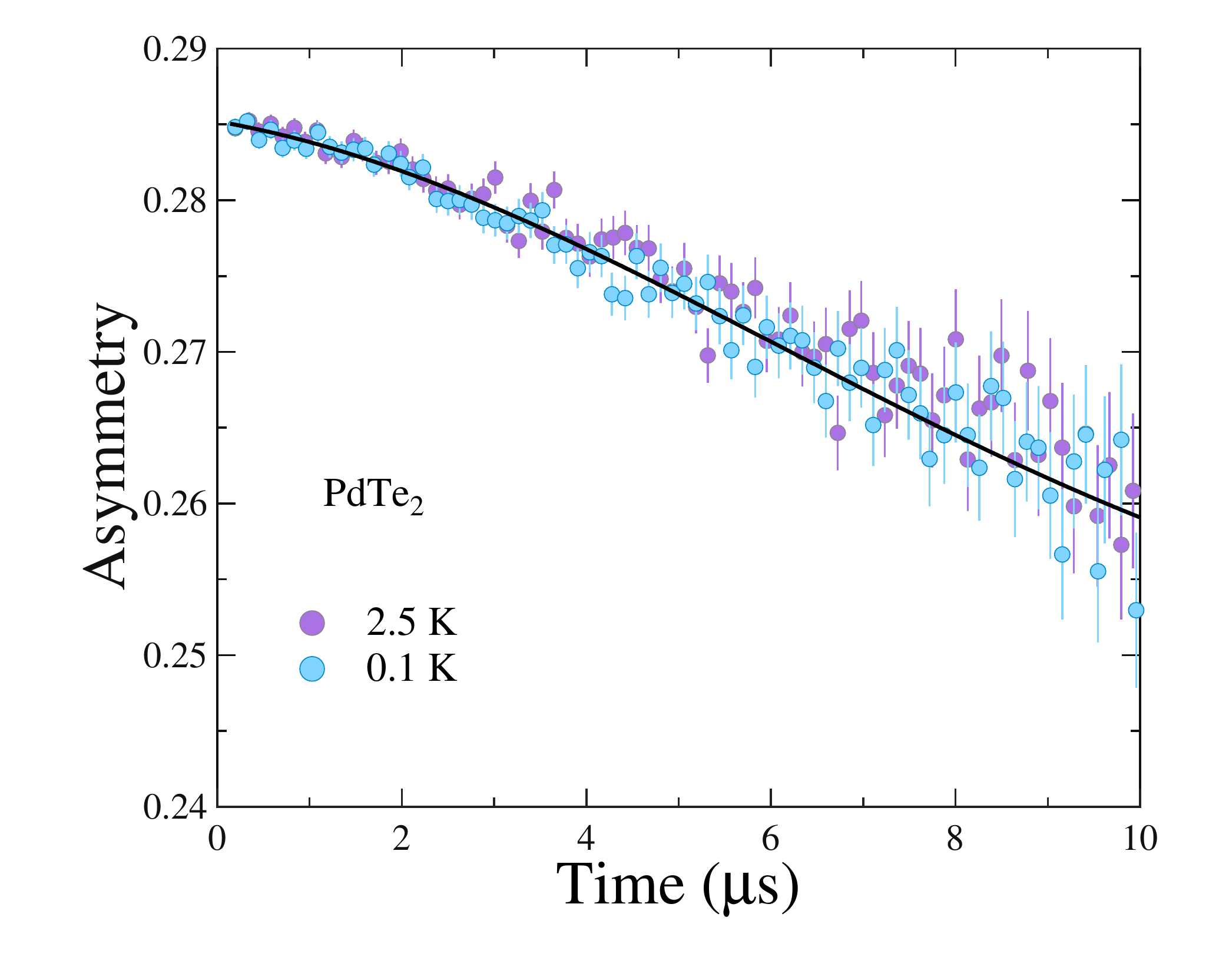}
\caption{\label{Fig1:Fig1} Representative ZF-$\mu$SR spectra collected below (0.1 K) and
above (2.5 K) the superconducting transition temperature. The solid
lines are the fits to Guassian Kubo-Toyabe (KT) function given in Eq. \eqref{eqn1:zf}.}
\end{figure}
ZF-$\mu$SR asymmetry spectra do not show any oscillatory signal which rules out the presence of large internal magnetic fields associated with long-range magnetic order. The time-dependent asymmetry spectra above and below $T_{c}$ do not show any discernible difference, which implies there is no additional relaxation signal in the superconducting state of PdTe$_{2}$. 
The ZF-$\mu$SR spectra are best described by the Gaussian Kubo-Toyabe (KT) function,
\begin{equation}
G_{\mathrm{KT}}(t) =  A_{1}\left[\frac{1}{3}+\frac{2}{3}(1-\sigma^{2}_{\mathrm{ZF}}t^{2})e^{-\sigma^{2}_{\mathrm{ZF}}t^2/2}\right]e^{-\Lambda t}+A_{\mathrm{BG}} ,
\label{eqn1:zf}
\end{equation} 
where $A_{1}$ is the initial asymmetry, $\sigma_{\mathrm{ZF}}$ denotes the relaxation due to static, randomly oriented local fields associated with the nuclear moments at the muon site. $A_{\mathrm{BG}}$ is the time-independent background contribution from the muons stopped in the sample holder whereas the exponential term, $\Lambda$, accounts for the presence of electronic relaxation channels.\\
\begin{figure}
\includegraphics[width=1.0\columnwidth]{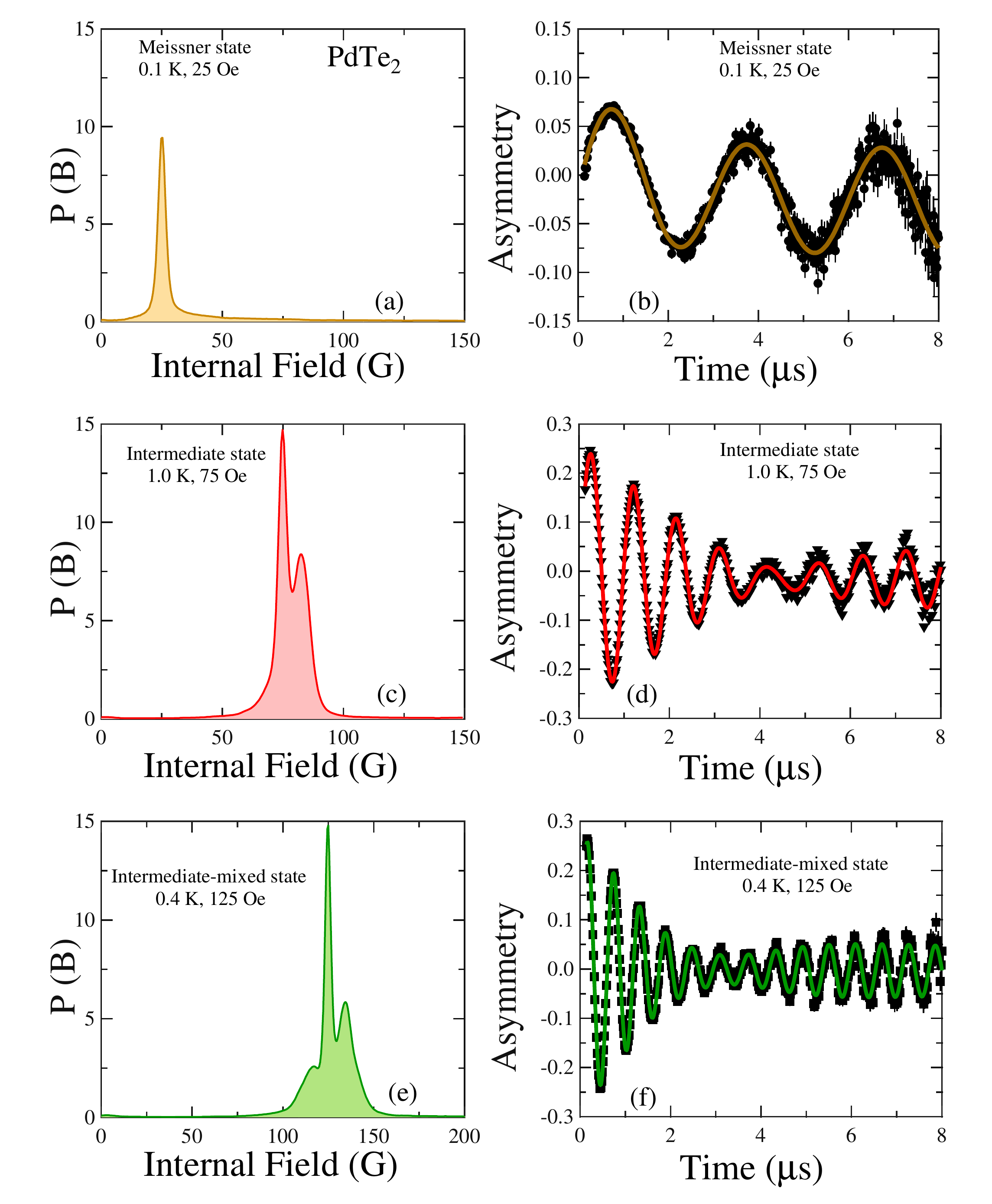}
\includegraphics[width=1.0\columnwidth]{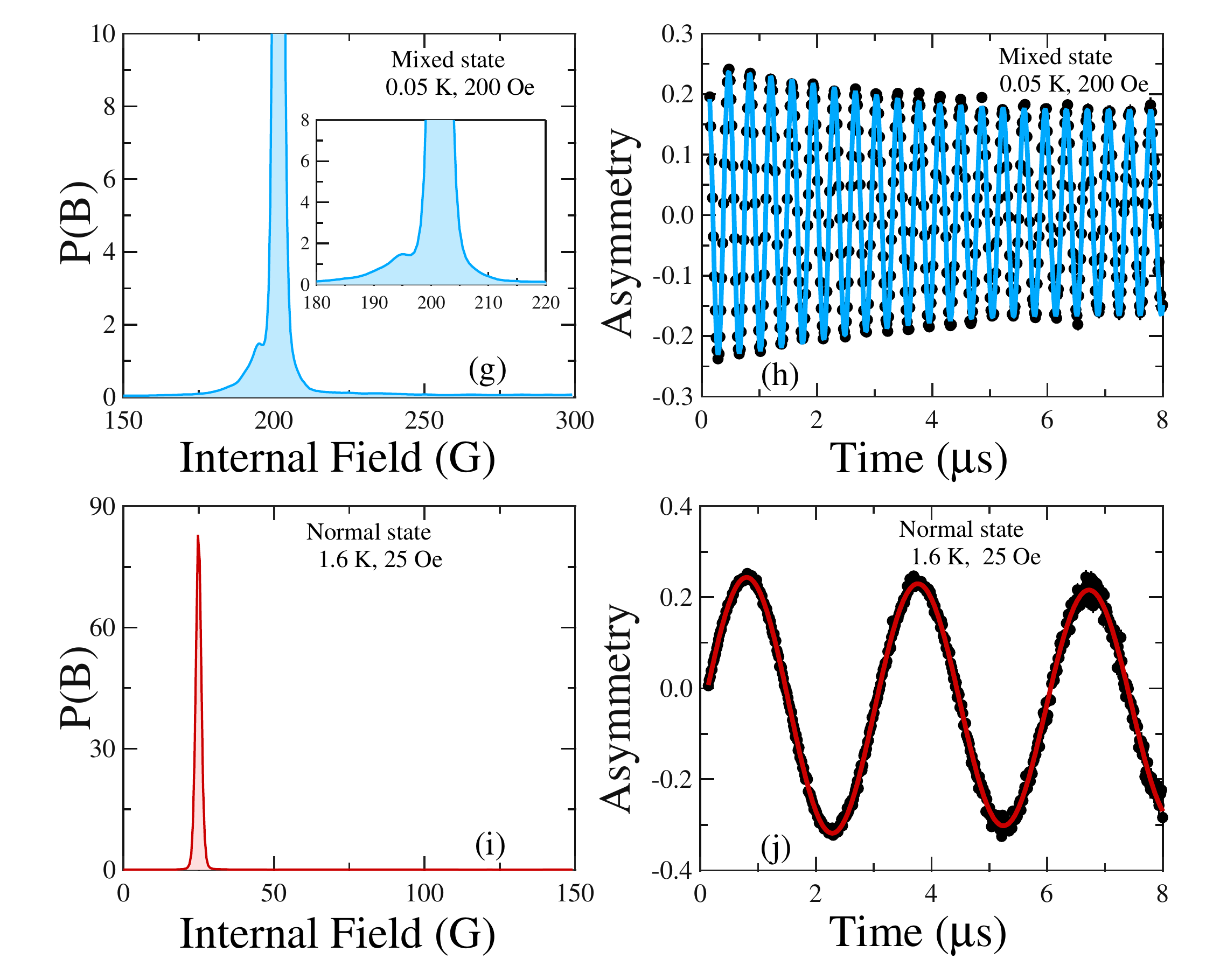}
\caption{\label{Fig3:Fig3}Field distribution of the local field probed by the muons, P(B), obtained by MaxEnt transformation of the TF-$\mu$SR time spectra at different temperatures and applied field. The figure illustrates
typical signal observed in the (a) Meissner, (c) Intermediate, (e) Intermediate-mixed state, (g) Mixed state, and (i) Normal state. (b), (d), (f), (h), and (j) show the TF-$\mu$SR time spectra for the corresponding states.}
\end{figure}
Materials exhibiting TRS breaking give rise to small spontaneous magnetic fields, which is readily detected by ZF-$\mu$SR as an increase in either $\sigma_{\mathrm{ZF}}$ or $\Lambda$. 
For temperatures below and above $T_{c}$ the relaxation rates $\sigma_{\mathrm{ZF}}$ and $\Lambda$ do not change within the resolution of the instrument ($\Delta$$\sigma_{\mathrm{ZF}}$ < 0.002 $\mu$$s^{-1}$, $\Delta$$\Lambda$ < 0.002 $\mu$$s^{-1}$). This indicates the absence of any spontaneous magnetic fields occurring in the
superconducting state, hence, there is no evidence of time-reversal symmetry
breaking in PdTe$_{2}$.\\
\begin{figure*}[t]
\includegraphics[width=2.0\columnwidth]{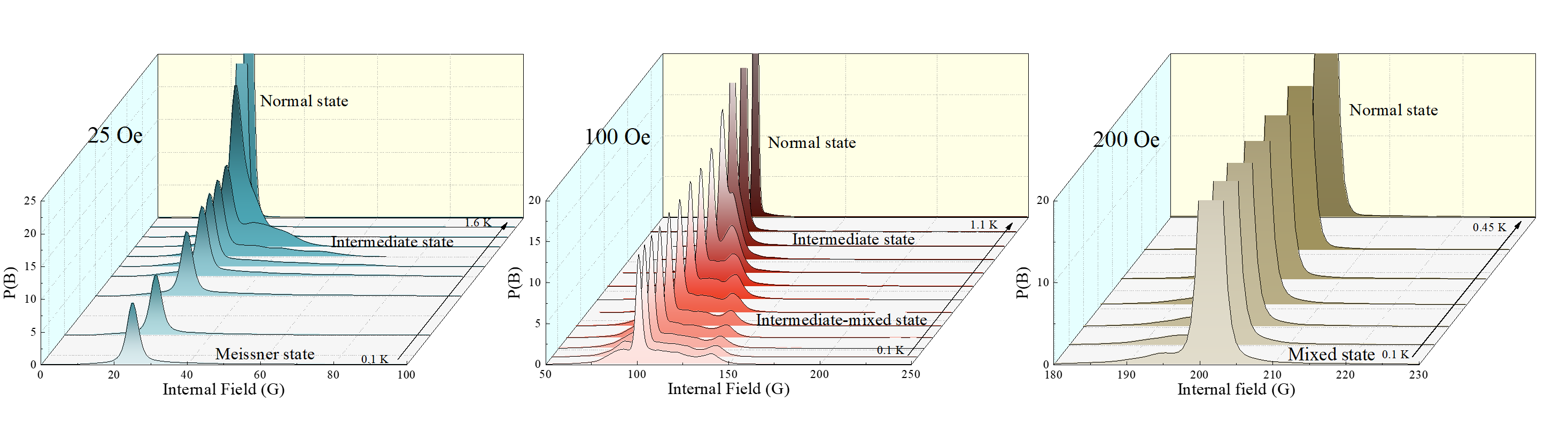}
\caption{\label{Fig5:Fig5}The figure illustrates the temperature dependence of the internal field distribution displaying a change from (a) Meissner to Intermediate to normal state at 25 Oe, (b) semi-
intermediate to intermediate state to normal state at 100 Oe, and (c) mixed state to normal state at 200 Oe.}
\end{figure*}
 The TF-$\mu$SR measurements were performed with the different applied magnetic fields up to 225 Oe to map out the full superconducting phase diagram of PdTe$_{2}$. As explained the TF-$\mu$SR precession signals were transformed into a probability field distribution using the maximum entropy (MaxEnt) algorithm \cite{BDR} to follow the evolution of different superconducting states in PdTe$_{2}$. All the TF-$\mu$SR data displayed in the present work were taken in the field cooled (FC) mode. As TF-$\mu$SR measurements analyze the entire volume of the material, the asymmetry spectra obtained directly depend on the nature of the superconducting state of the sample. For example, the magnetic moment of a muon landing within a type-I superconductor will either be stationary in the superconducting regions or will precess in a field very close to the thermodynamic critical field, $H_{c}$, in the normal regions. Meanwhile, for a type-II superconductor in the mixed state, the muons will precess with a field component that is lower than the applied field. Therefore, the probability of muon experiencing the internal field is equal to the volume fraction of the sample in the respective states. The Fourier transform of the TF-$\mu$SR spectra (MaxEnt transformation) gives the probability distribution of internal field, $H_{int}$, in both the intermediate state of
a type-I superconductor and the mixed state of a type-II superconductor.

 Figure \ref{Fig3:Fig3} (a) and (b) show the MaxEnt transformation and asymmetry spectra at $T$ = 0.1 K, and $H_{app}$ = 25 Oe. At this temperature and field, PdTe$_{2}$ is in Meissner state. The TF asymmetry spectrum in Fig. \ref{Fig3:Fig3} (b) shows large oscillations as expected for muons stopping in the silver holder and only see the applied field, $H_{app}$, giving a background signal. In the MaxEnt data shown in Fig. \ref{Fig3:Fig3} (a) this is well represented by peak near $B_{int}$ = 25 G. The absence of any additional peaks (> $H_{app}$ and < $H_{app}$) implies that the magnetic field is completely expelled from the body of the superconductor. Reportedly, it should have additional contributions coming from the fraction of muons stopping in the sample. These muons precess with the local field of the sample, which is generally contributed by the nuclear moments of the constituent elements and manifested in the MaxEnt data by the peak near low fields ($B_{int}$ $\approx$ 0 G). Surprisingly, we did not observe any additional peak near $B_{int}$ $\approx$ 0 G. This is completely understandable if we consider the nuclear moment values for Pd and Te, which is very small and makes it very difficult to extract the nuclear field contribution from MaxEnt transformation. Nevertheless, the TF-$\mu$SR spectra in the Meissner state ($T$ = 0.1 K, $H_{app}$ = 25 Oe) displays an additional nonoscillating component due to random nuclear dipole moments of the Pd/Te atoms which can be fitted using the KT relaxation function. The relaxation rate of this nonoscillating component is equal to the relaxation rate of measurements performed in zero applied field, and so this component may be identified with superconducting regions of the sample.\\  
TF-$\mu$SR spectra further show a considerable reduction in the initial asymmetry. The loss of initial asymmetry as observed in TF-$\mu$SR spectra of PdTe$_{2}$ is similar to that observed in type-I superconductors LaRhSi$_{3}$ \cite{VKA1}, LaNiSn \cite{LNS} and the recently published BeAu \cite{BA1,BA2}. The missing asymmetry behavior is due to the detectors in the TF geometry unable to detect the positrons emitted in the muon spin polarization direction when the sample is in the Meissner state \cite{BA1,BA2}. In other words, muons landing in the superconducting regions of the sample in the Meissner state will be stationary and less likely to be detected. At temperature near $T_{c}$ where the sample is in normal state, the initial asymmetry recovers to its full maximum value [see \figref{Fig3:Fig3} (j)].

 Figure \ref{Fig3:Fig3} (c) and 3 (d) shows the typical MaxEnt results of magnetic field distribution and TF-$\mu$SR in the intermediate state of PdTe$_{2}$ at $T$ = 1.0 K and $H_{app}$ = 75 Oe. In the intermediate state normal and superconducting regions coexist, therefore, the total $\mu$SR asymmetry is the fraction of the signal coming from each respective region. The data is analyzed using the following function:
\begin{eqnarray}
G_{\mathrm{z}}(t)=  G_{KT}(t)+\sum_{i=1}^N A_{i}\exp\left(-\frac{1}{2}\sigma_i^2t^2\right)\cos(\omega_it+\phi),
\label{eqn2:TF}
\end{eqnarray}
where G$_{KT}$(t) is Eq. \eqref{eqn1:zf}, $A_{i}$ is the initial asymmetry, $\sigma_i$ is the Gaussian relaxation rate, $\phi$ is the common phase offset, and $\omega_i$ is the frequency of the muon spin precessional signal of the respective components. We found that the asymmetry spectra can best be described by three oscillating functions (N=3). In these fits, $\omega_{1}$ = $\omega_{2}$ and $\sigma_1$ = 0 means that there are two components ($A_{1}$ and $A_{2}$) precessing about the applied field, one with a decaying component and one not. Nondecaying component corresponds to the background term arising from those muons stopping in the silver sample holder. A fit to Eq. \eqref{eqn2:TF} yields $A_{2}$ = 0.0872 and muon precessional frequency of 1.016 MHz and 1.115 MHz in the superconducting state. Total asymmetry from the sample in the normal state of 0.278 means the nonsuperconducting volume fraction is estimated to be 31.3 $\%$. The muon precession frequencies is related to the local field strength by
\begin{equation}
f= \frac{\gamma_{\mu}}{2\pi}B ,
\label{eqn3:LS}
\end{equation}
where $\gamma_{\mu}/2\pi$ = 135.5 MHz/T is the muon gyromagnetic ratio. This implies internal magnetic fields is 75 G and 83 G. The internal field $B_{int}$ $\approx$ 75 G, which is exactly equal to the applied field gives the background signal, whereas $B_{int}$ $\approx$ 83 G can be taken as an estimate of the critical field, $H_{c}$, coming from the normal regions in the  intermediate state. The presence of an internal field greater than the applied field is strong evidence of type-I superconductivity in the compound. It is worth mentioning that the intermediate state of a type-I superconductor is induced due to non-zero demagnetization effects. Demagnetizing effects cause some regions of the sample experiencing a field greater than the $H_{c}$, even if the applied magnetic field, $H_{app}$, is considerably less than the critical field $H_{c}$. In this situation, the superconductor will have a complicated structure where the normal-state domains coexist with the superconducting domains in space. Muons implanted in these normal regions of the intermediate state will precess at a frequency corresponding to the field equal to $H_{c}$, whereas the muons landing in the superconducting regions will be stationary and only be affected by the nuclear moments. The muons stopping in the silver holder will precess about the applied magnetic field. In the MaxEnt data shown in \figref{Fig3:Fig3} (c), the intermediate state is well demonstrated by the two sharp peaks ($B_{int}$ $\approx$ 75 G and $B_{int}$ $\approx$ 83 G).

 MaxEnt visualization and the corresponding time-dependent asymmetry spectra in \figref{Fig3:Fig3} (e) and 3(f) exhibit regions with the coexistence of intermediate and mixed state. At $T$ = 0.4 K and $H_{app}$ = 125 Oe [see \figref{Fig3:Fig3} (e)], the MaxEnt data show characteristic features of mixed and intermediate state. Notably, for a type-II superconductor in the mixed state, a field component at a lower value than the applied field is expected due to the establishment of the flux line lattice. The muons will precess at a frequency $\omega$ = $\gamma_{\mu}$$B_{int}$ where $B_{int}$ is an asymmetric distribution, the peak of which falls below $H_{app}$. This is clearly present at $B_{int}$ = 115 G, indicating type-II superconductivity and strong evidence for the presence of vortices in the superconducting state of PdTe$_{2}$. In previous reports, the formation of the vortex core is seen in the superconducting state of PdTe$_{2}$. However, the formation of FLL region was not clearly resolved \cite{OJ}. In our measurements, we found the formation of FLL around $H_{app}$ = 75 Oe which continues to appear up to the magnetic field of 200 Oe [see \figref{Fig7:Fig7}].\\ Meanwhile, a peak near $B_{int}$ = 135 G > $H_{app}$, demonstrates the intermediate state of type-I superconductivity in PdTe$_{2}$. In addition, fitting the time spectra data [see \figref{Fig3:Fig3} (f)] we determined three frequencies 1.55 MHz (115 G), 1.7 MHz (125 G) and 1.83 MHz (135 G), corresponding to magnetic fields due to the formation of the FLL, background applied field, and the intermediate state, respectively. This study provides direct proof for the coexistence of the type-I/type-II superconductivity in this compound. It is argued that the observed admixture of type-I and type-II superconductivity in PdTe$_{2}$ arises due to inhomogeneous electron mean free path, $l$, on the surface of PdTe$_{2}$. The inhomogenous mean free path of the electrons modifies the coherence length $\xi$, and penetration depth $\lambda$, in the superconducting state which subsequently pushes the system to the type-II limit \cite{ASS,TLL}. Interestingly, similar evidence of type-I and type-II superconductivity and their coexistence was observed in ZrB$_{12}$ \cite{ZrB}.\\Another representative spectrum is shown in \figref{Fig3:Fig3} (h) at T = 0.05 K and $H_{app}$ = 200 Oe, where an increase in the depolarization rate is observed due to the formation of inhomogeneous field distribution. Fits to the time-dependent spectra data reveal two frequencies of 2.71 MHz and 2.62 MHz corresponding to magnetic fields of 200 and 194 Oe, respectively. Again the high-frequency signal is the background signal whereas the other signal with a lower frequency is due to the formation of the FLL in the mixed state. In the maximum entropy data [see \figref{Fig3:Fig3} (g)], a Gaussian distribution of fields due to the FLL is observed below the applied field and this indicates that the sample is in the mixed state of a type-II superconductor. At $H_{app}$ = 200 Oe, the regions with type-I superconductivity become normal, while the type-II superconducting region persists. This indicates that the points where superconductivity is destroyed due to applied magnetic field corresponding to type-I regions, are different from the points where we have type-II superconductivity.\\ Finally, at an applied field of 225 Oe and $T$ = 0.1 K, PdTe$_{2}$ returns to the normal state. Here, the field penetrates the bulk of the sample completely, and we see homogeneous field distribution in the TF-$\mu$SR time spectra [see Fig. \ref{Fig3:Fig3} (j)], corresponding to a single peak at the applied field position in the MaxEnt data [see Fig. \ref{Fig3:Fig3} (i)]. This suggests that the critical field from the $\mu$SR measurements is $H_{c}$ $\approx$ 225 Oe. Intriguingly, the results presented in Ref. \cite{ASS,TLL} show a spatial distribution of critical fields which extend up to 400 Oe-600 Oe. Here, the points on the crystal showing lower critical field are type-I while the points showing high critical fields resemble type-II superconductors. This discrepancy of results could be because the critical fields determined in Ref. \cite{ASS,TLL} is from the surface sensitive techniques where the defects on the surface of the crystal play an essential role. On the contrary, $\mu$SR is more closely related to the bulk measurements where muon implants deep inside the sample and measure the internal field distribution.\\ 
\begin{figure}
\includegraphics[width=1.0\columnwidth]{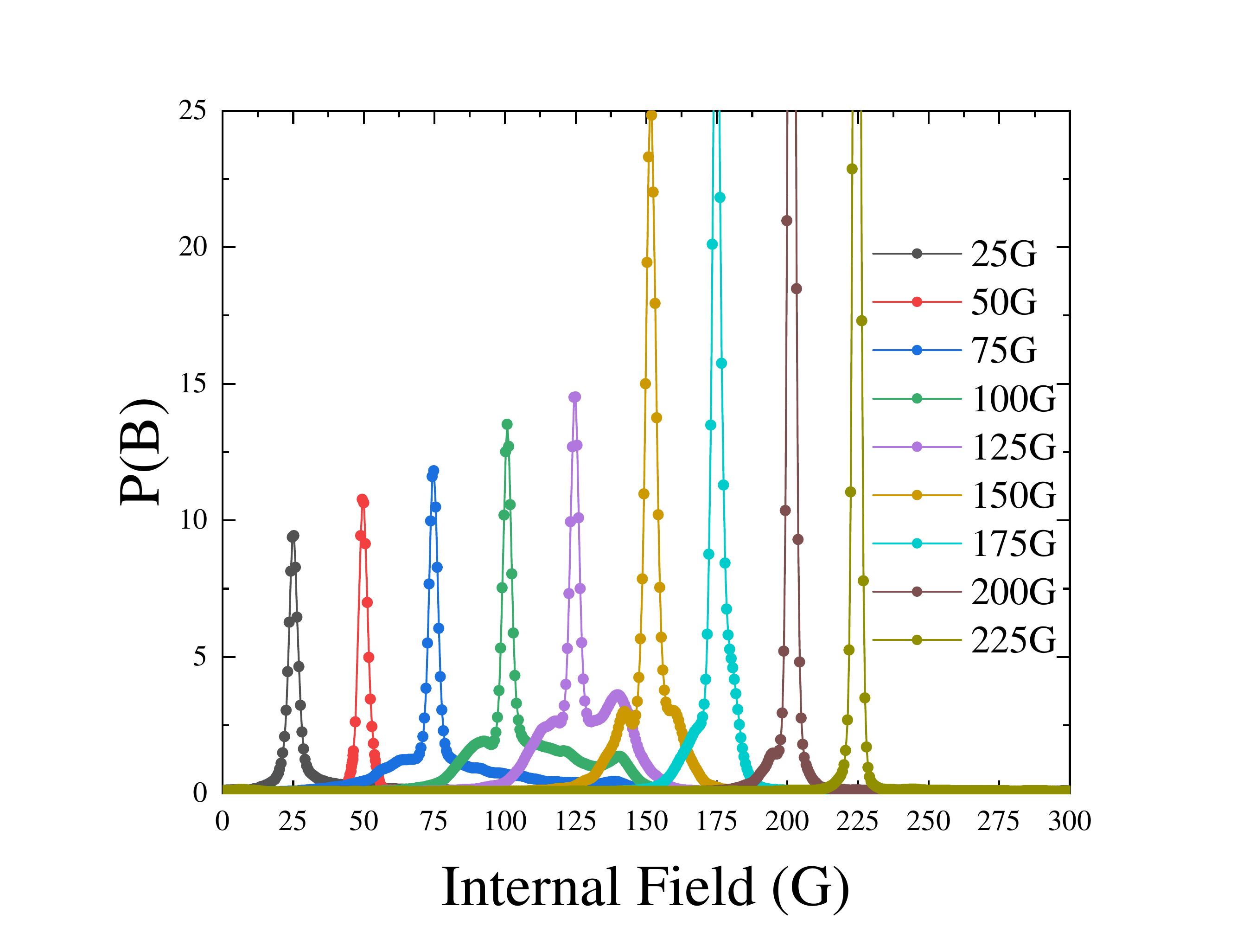}
\caption{\label{Fig6:Fig6}  Field distribution of the internal field probed by the muons, P(B),  at different fields between 25 Oe $\le$ H $\le$ 225 Oe at $T$ = 0.1 K.}
\end{figure}
\begin{figure}
\includegraphics[width=1.0\columnwidth]{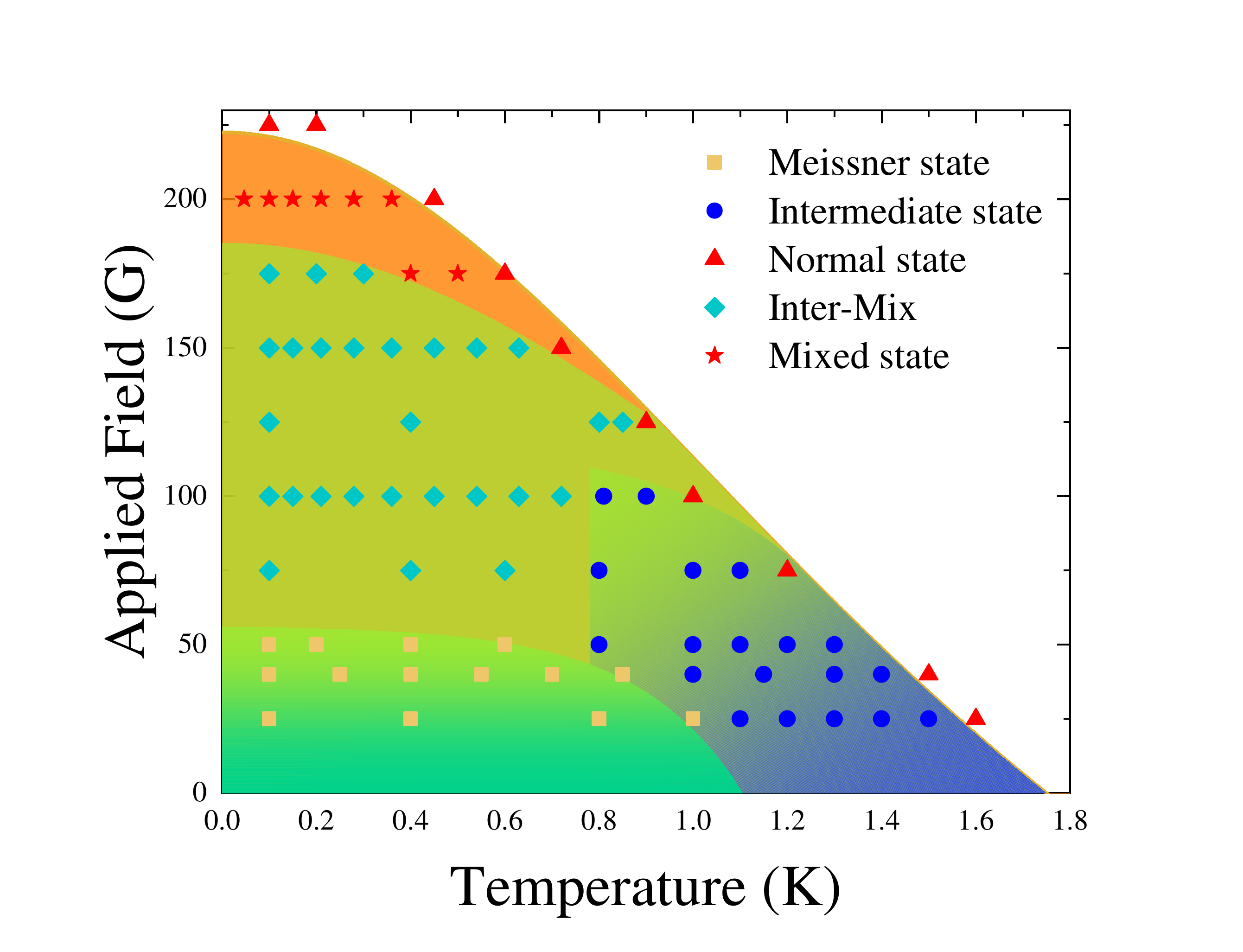}
\caption{\label{Fig7:Fig7}  The phase diagram. Different superconducting phases of PdTe$_{2}$ determined from the muon spin rotation experiments. The data comes directly from the MaxEnt analysis of the TF-$\mu$SR spectra by measuring the peak positions of the internal field  distribution. The phase boundaries were determined by the fitting of the time dependent signal.}
\end{figure}
To draw a full superconducting phase diagram of PdTe$_{2}$ experimentally, we have investigated the local field distribution, P(B), as a function of temperature. Maximum entropy data was determined in a temperature range between 0.1 K $\le$ T $\le$ 1.6 K, as shown in \figref{Fig5:Fig5}. Figure \ref{Fig5:Fig5} (a) shows the temperature dependence of the internal field distribution at 25 Oe, displaying a change from the Meissner to the intermediate to the normal state with increasing temperature. At T = 0.1 K, a single peak appears in the maximum entropy data, P(B), at $B_{int}$ $\simeq$ $H_{app}$ $\simeq$ 25 G, reminiscent of the Meissner state. As explained earlier, the peak near $B_{int}$ $\approx$ 0 G is absent due to the negligibly small values for nuclear moments of Pd/Te atoms. As we increase the temperature, the sample remains in the Meissner state until it reaches $T$ = 1.0 K. For temperatures above 1.0 K, the system shifts to the intermediate state. Finally, at $T$ = 1.6 K, the sample enters the normal state.  Similarly, \figref{Fig5:Fig5} (b) exhibits a shift from the intermediate-mixed state to the intermediate state to the normal state with increasing temperature at 100 Oe. At $T $= 0.1 K, three different peaks appear in the P(B) spectra at $B_{int}$ $\approx$ 94 G, $B_{int}$ $\approx$ $H_{app}$ $\approx$ 100 G and $B_{int}$ $\approx$ 135 G. These peak positions indicate a mixed state for a type-II superconductor, background signal coming from the applied field, and intermediate state in type-I superconductor. With an increase in temperature, the peak position for $H_{c}$ moves closer to $H_{app}$ with a subsequent increase in the magnitude. The observed behavior is due to an increase in the volume of the normal core regions in the intermediate state. On the other hand, the peak due to the mixed state is suppressed in magnitude as more field lines start to penetrate the sample. Eventually, when the transverse applied field is increased to 200 Oe [see \figref{Fig5:Fig5} (c)], the intermediate state due to type-I superconductivity is completely suppressed. While the peak around the applied field broadens and an additional shoulder in the distribution is observed at lower fields indicating type-II superconductivity. This is the field distribution of the FLL in the mixed state, where most of the contribution of P(B) is at fields less than the applied field. This state persists up to $T$ < 0.45 K, above which the sample enters the normal state.
 
 To further investigate the behavior of superconducting phase in PdTe$_{2}$, we plotted the MaxEnt transformation, P(B), as a function of different fields between 25 Oe $\le$ $H_{app}$ $\le$ 225 Oe. Figure \ref{Fig6:Fig6} shows the internal field distribution at $T$ = 0.1 K, clearly demonstrating the change from the Meissner state (25 Oe $\le$ $H_{app}$ $\le$ 50 Oe) to the intermediate-mixed state (75 Oe $\le$ $H_{app}$ $\le$ 175 Oe) to the mixed state ($H_{app}$ = 200 Oe) with increasing magnetic field. At the intermediate-mixed state for $H_{app}$ $>$ 50 Oe, P(B) show additional internal peaks at $B_{int}$ $<$ $H_{app}$ and $B_{int}$ $>$ $H_{app}$ corresponding to the FLL and intermediate state in the sample. A close inspection of \figref{Fig6:Fig6} reveals that as the magnetic field is increased, type-I superconducting features in PdTe$_{2}$ grow more distinctly as compared to the mixed state in the type-II superconducting regions. For example, in the magnetic field range between 25 Oe and 125 Oe, the peak due to the intermediate state becomes more pronounced compared to the peak due to the mixed state. However, as the field is increased to $H_{app}$ = 150 Oe, the magnitude of peaks resulting from both the states becomes comparable, suggesting equivalent contributions. For $H_{app}$ $>$ 150 Oe, the peaks due to the intermediate state is suppressed as a result of an increase in normal core regions while the points corresponding to the FLL region disappear gradually as in type-II superconductors. At $H_{app}$ = 200 Oe, the superconductivity in type-I regions is destroyed whereas we can still observe a shoulder at lower fields ( $<$ $H_{app}$) due to the mixed state in the system. In the end, at $H_{app}$ = 225 Oe the system goes to the normal state which is apparent from the presence of only one peak in the P(B) graph.
 
 Through systematic temperature and magnetic field dependent $\mu$SR experiments, we mapped out the superconducting phase diagram for PdTe$_{2}$. By measuring the peak position of the local field distribution, $B_{int}$, we find evidence of the Meissner, intermediate, and mixed states in PdTe$_{2}$. The obtained phase diagram is shown in \figref{Fig7:Fig7}. The existence of three separate superconducting phases is color-coded in green, blue, and orange regions, respectively. While the mixed state shows the internal field distribution due to the formation of the FLL, carrying the signature of type-II superconductivity, the intermediate state is a characteristic feature of type-I superconductivity. The Meissner state is common to both type-I and type-II superconductors. We also see regions where different states coexist. We ascribe this region as a semi-intermediate state where the mixed state coexists with the intermediate state shown by light green area in the middle. In the H-T phase diagram [see \figref{Fig7:Fig7}], the Meissner state is denoted by the yellow square markers, whereas the intermediate state appears at high temperatures is displayed by blue circles. With an increase in the magnetic field, the semi-intermediate comes into the picture, represented by cyan diamond markers. A further increase in the field leads to a mixed state highlighted by red star markers. These results indicate that PdTe$_{2}$ is a marginal superconductor between type-I and type-II phases. Furthermore, the variation of the local field distribution at different magnetic fields and temperature demonstrate that it is possible to drive the system into various superconducting states with the help of external parameters, such as field and temperature.
 
 Prior to this work, the status of the superconducting state of PdTe$_{2}$ seemed rather ambiguous. The current work which samples the local field distribution, P(B), decisively demonstrates that the superconducting state of PdTe$_{2}$ resides near the critical point between its type-I to type-II phases. Indeed, our $\mu$SR measurements illustrate that for a field between 75 Oe and 175 Oe, there exists an overlap of type-I and type-II superconducting phases. We recall that the ratio of coherence length $\xi$, and penetration depth $\lambda$, defines the Ginzburg-Landau parameter, $\kappa$ = $\frac{\lambda}{\xi}$. Superconductors with $\kappa$ < 1/$\sqrt{2}$ classified as type-I, whereas $\kappa$ > 1/$\sqrt{2}$ as type-II. Recently, Tian Le \textit{et al.} plotted the mean free path, $l$, in accordance with the Ginzburg-Landau parameter, $\kappa$, and demonstrated that a reduced $l$ could cause an enhanced $\kappa$ \cite{TLL}. In fact, mean free path of $l$ $\le$ 200 nm on some surface regions can cause an enhanced $\kappa$ (> 1/$\sqrt{2}$) and thus transform it into a type-II superconductor. It is interesting to note that our $\mu$SR measurements reveal that it is possible to shift to different states as a function of temperature and field which reversibly produces both type-I/type-II and admixed state, respectively. It possibly indicates the temperature and field dependence of $\kappa$, which is related to the observed phase transition. It is evident that if the $\kappa$ value of a superconductor is very close to 1/$\sqrt{2}$ and also dependant on temperature and magnetic field, all of these exotic states may coexist in the same phase diagram, which is shown in \figref{Fig7:Fig7}. 
\section{Conclusion}
 In summary, we have carried out the $\mu$SR experiments on the single crytals of PdTe$_{2}$. Zero-field measurements show no evidence for a time-reversal symmetry-breaking field in the superconducting state. Using temperature and field dependent TF-$\mu$SR spectra, we have drawn the superconducting phase diagram of PdTe$_{2}$. The local field distribution for different applied fields and temperatures suggests a mixture of type-I and type-II superconductivity with the critical field, $H_{c}$ $\approx$ 225 Oe. The inhomogeneity on the sample crystal surface causes a spatial distribution of critical fields, which in some cases ranges from 220 Oe to 4 T \cite{SD}. From a statistical point of view, the critical field is found to be around 250 Oe. This is in close agreement with the critical field measured by the heat capacity, transport measurements and the value estimated from our $\mu$SR experiments. It is important to note that the higher critical fields were obtained from the tunneling and spectroscopy measurements, although there is no signature of magnetic vortices in the presence of the magnetic field \cite{SD,ASS,TLL}. In our $\mu$SR measurements, however, we can see the signatures of FLL regions in field as small as 75 Oe. We speculate that the lower critical field in $\mu$SR measurements is observed due to the bulk nature of measurenet technique. Therefore, in future we would like to extend our studies using low energy muon beams \cite{EMR} to discriminate between near-surface and bulk superconductivities. It would be of considerable interest if the surface superconductivity were influenced by multi-band effects and appreciably different from that of the bulk.\\
 
$\textit{Note added}$. Recently, we became aware of Ref. \cite{HLJ}, which reports similar results on PdTe$_{2}$. Interestingly, in Ref. \cite{HLJ} type-I behaviour is predicted by the presence of intermediate state.  From our measurements we were able to detect the signatures of both type-I and type-II superconducting behavior.
\section{Acknowledgments}

We thank ISIS, STFC, UK for the Newton funding and beamtime to conduct the $\mu$SR experiments. Work at CAU was supported by the Rare Isotope Science Project of Institute for Basic Science funded by Ministry of Science and ICT and NRF of Korea (2013M7A1A1075764) .


\begin{thebibliography}{References}

\bibitem{AP} A. P. Schnyder, S. Ryu, A. Furusaki, and A. W. W. Ludwig,
Phys. Rev. B 78, 195125 (2008).

\bibitem{XL1} X.-L. Qi, T. L. Hughes, S. Raghu, S.-C. Zhang, Phys. Rev. Lett.
102, 187001 (2009).

\bibitem{XL2} X.-L. Qi and S.-C. Zhang, Rev. Mod. Phys. 83, 1057 (2011).

\bibitem{MS} M. Sato and S. Fujimoto, Phys. Rev. B 79, 094504 (2009).

\bibitem{ZSM} Z. Sun, M. Enayat, A. Maldonado, C. Lithgow, E. Yelland,
D. C. Peets, A. Yaresko, A. P. Schnyder, and P. Wahl, Nat. Commun. 6, 6633 (2015).

\bibitem{NR} N. Read and D. Green, Phys. Rev. B. 61, 10267 (2000).

\bibitem{AK} A. Kitaev, Phys. Usp. 44, 131 (2001).

\bibitem{LFC} L. Fu and C. L. Kane, Phys. Rev. Lett. 100, 096407 (2008).

\bibitem{MZH} M. Z. Hasan and C. L. Kane, Rev. Mod. Phys. 82, 3045 (2010).

\bibitem{ZL} Z. Liu, X. Yao, J. Shao, M. Zuo, L. Pi, S. Tan, C. Zhang, and Y. Zhang, 
J. Am. Chem. Soc. 137, 33 (2015).

\bibitem{MPS1} M. P. Smylie, K. Willa, H. Claus, A. Snezhko, I. Martin, W.-K.
Kwok, Y. Qiu, Y. S. Hor, E. Bokari, P. Niraula, A. Kayani, V.
Mishra, and U. Welp, Phys. Rev. B 96, 115145 (2017).

\bibitem{MPS2} M. P. Smylie, K. Willa, H. Claus, A. E. Koshelev, K. W. Song,
W.-K. Kwok, Z. Islam, G. D. Gu, J. A. Schneeloch, R. D. Zhong,
and U. Welp, Sci. Rep. 8, 7666(2018).

\bibitem{YSH} Y. S. Hor, A. J. Williams, J. G. Checkelsky, P. Roushan, J. Seo,
Q. Xu, H.W. Zandbergen, A. Yazdani, N. P. Ong, and R. J. Cava,
Phys. Rev. Lett. 104, 057001 (2010).

\bibitem{YQ} Y. Qiu, K. N. Sanders, J. Dai, J. E. Medvedeva, W. Wu, P. Ghaemi, T. Vojta, and Y. S. Hor, arXiv:1512.03519.

\bibitem{ZWANG} Z. Wang, A. A. Taskin, T. Frolich, M. Braden, and Y. Ando, Chem. Mat. 28, 779 (2016)

\bibitem{CB} C. Beenakker and L. Kouwenhoven, Nat. Phys. 12, 618 (2016).

\bibitem{VS} V. S. Pribiag, A. J. A. Beukman, F. Qu, M. C. Cassidy, C. Charpentier, W. Wegscheider, and L. P. Kouwenhoven, Nat. Nanotechnol. 10, 593 (2015).

\bibitem{PZ} P. Zareapour, A. Hayat, S. Y. F. Zhao, M. Kreshchuk, A. Jain, D. C.
Kwok, N. Lee, S.-W. Cheong, Z. Xu, A. Yang, G. D. Gu, S. Jia, R. J.
Cava, and K. S. Burch, Nat. Commun. 3, 1056 (2012).

\bibitem{JLZ} J. L. Zhang, S. J. Zhang, H. M. Weng, W. Zhang, L. X. Yang,
Q. Q. Liu, S. M. Feng, X. C. Wang, R. C. Yu, L. Z. Cao, L. Wang, W. G. Yang, H. Z. Liu, W. Y. Zhao, S. C. Zhang, X. Dai, Z. Fang, and C. Q. Jin, Proc. Natl. Acad. Sci. USA 108, 24 (2011).

\bibitem{LAA} L. Aggarwal, A. Gaurav, G. S. Thakur, Z. Haque, A. K. Ganguli,
and G. Sheet, Nat. Mater. 15, 32 (2015).

\bibitem{HWH} H. Wang, H. Wang, H. Liu, H. Lu, W. Yang, S. Jia, X. J. Liu,
X. C. Xie, J. Wei, and J. Wang, Nat. Mater. 15, 38 (2015).

\bibitem{SDL} S. Das, L. Aggarwal, S. Roychowdhury, M. Aslam, S. Gayen,
K. Biswas, and G. Sheet, Appl. Phys. Lett. 109, 132601
(2016).

\bibitem{AA} A. A. Soluyanov, D. Gresch, Z. Wang, Q. Wu, M. Troyer, X. Dai, and B. Bernevig, Nature 527, 495-498 (2015).

\bibitem{HH} H. Huang, S. Zhou, and W. Duan, Phys. Rev. B 94, 121117(R) (2016).

\bibitem{MY} M. Yan, H. Huang, K. Zhang, E. Wang, W. Yao, K. Deng, G. Wan, H. Zhang, M. Arita, H. Yang, Z. Sun, H. Yao, Y. Wu, S. Fan, W. Duan, and S. Zhou, Nature Comm. 8, 257 (2017).

\bibitem{MSB} M. S. Bahramy, O. J. Clark, B. J. Yang, J. Feng, L. Bawden, J. M. Riley, I. Markovic, F. Mazzola, V. Sunko, D. Biswas, S. P. Cooil, M. Jorge, J. W. Wells, M. Leandersson, T. Balasubramanian, J. Fujii, I. Vobornik, J. E. Rault, T. K. Kim, M. Hoesch, K. Okawa, M. Asakawa, T. Sasagawa, T. Eknapakul, W. Meevasana, and P. D. C. King Nature Mat. 17, 21 (2018).

\bibitem{YL} L. Yan, Z. Jian-Zhou, Y. Li, L. Cheng-Tian, H. Cheng, L. De-Fa,
P. Ying-Ying, X. Zhuo-Jin, H. Jun-Feng, C. Chao-Yu et al., Chin.
Phys. B 24, 067401 (2015).

\bibitem{FF} F. Fei, X. Bo, R. Wang, B. Wu, J. Jiang, D. Fu, M. Gao, H. Zheng, Y. Chen, X. Wang, H. Bu, F. Song, X. Wan, B. Wang, and G. Wang, Phys. Rev. B 96, 041201 (2017).

\bibitem{HN}  H. Noh, J. Jeong, E.-J. Cho, K. Kim, B. I. Min, and B.-G. Park,
Phys. Rev. Lett. 119, 016401 (2017).

\bibitem{OJ} O. J. Clark, M. J. Neat, K. Okawa, L. Bawden, I. Markovi$\acute{c}$, F. Mazzola, J. Feng, V. Sunko, J. M. Riley, W. Meevasana, J. Fujii, I. Vobornik, T. K. Kim, M. Hoesch, T. Sasagawa, P. Wahl, M. S. Bahramy, and P. D. C. King, Phys. Rev. Lett. 120, 156401 (2018).

\bibitem{JG} J. Guggenheim, F. Hulliger, and J. Müller, Helv. Phys. Acta 34,
408 (1961).

\bibitem{FJ} F. Jellinek, Arkiv. Kemi 20, 447 (1963).

\bibitem{HLC} H. Leng, C. Paulsen, Y. K. Huang, and A. de Visser, Phys. Rev.
B 96, 220506 (2017).

\bibitem{DS} D. Saint-James and P. G. Gennes, Phys. Lett. 7, 306 (1963).

\bibitem{DS1} D. Saint-James, E. J. Thomas, and G. Sarma, \textit{Type II Superconductivity} (Pergamon, Oxford, UK, 1969)






\bibitem{AY} Amit and Y. Singh, Phys. Rev. B 97, 054515 (2018).

\bibitem{MVS} M. V. Salis, P. Rodi`ere, H. Leng, Y. K. Huang, and A. de
Visser, J. Phys: Condens. Matter 30, 505602 (2018).

\bibitem{STN} S. Teknowijoyo, N. H. Jo, M. S. Scheurer, M. A. Tanatar, K. Cho, S. L. Bu$\acute{d}$ko, P. P. Orth, P. C. Canfield, and R. Prozorov, Phys. Rev. B 98, 024508 (2018).

\bibitem{SD} S. Das, Amit, A. Sirohi, L. Yadav, S. Gayen, Y. Singh, and Go. Sheet Phys. Rev. B 97, 014523 (2018).

\bibitem{ASS} A. Sirohi, S. Das, P. Adhikary, R. R. Chowdhury, A. Vashist, Y. Singh, S. Gayen, T. Das and Go. Sheet, J. Phys.: Condens. Matter 31, 085701, (2019).

\bibitem{TLL} T. Le, L. Yin, Z. Feng, Q. Huang, L. Che, J. Li, Y. Shi, and X. Lu Phys. Rev. B 99, 180504(R) (2019).

\bibitem{VSE} V. S. Egorov, G. Solt, C. Baines, D. Herlach, and U. Zimmermann, Phys. Rev. B 64, 024524 (2001).

\bibitem{BA1} D. Singh, A. D. Hillier, and R. P. Singh, Phys. Rev. B 99,
134509 (2019).

\bibitem{BA2} J. Beare, M. Nugent, M. N. Wilson, Y. Cai, T. J. S. Munsie, A. Amon, A. Leithe-Jasper, Z. Gong, S. L. Guo, Z. Guguchia, Y. Grin, Y. J. Uemura, E. Svanidze, G. M. Luke, Phys. Rev. B 99, 134510 (2019).

\bibitem{LT} L. Thomassen, Z. Phys. Chem. B 2, 349 (1929).

\bibitem{SLL} S. L. Lee, S. H. Kilcoyne, and R. Cywinski, Muon Science: Muons in Physics, Chemistry and Materials (SUSSP Publications and IOP Publishing, Bristol, 1999).

\bibitem{BDR} B. D. Rainford and G. J. Daniell, Hyperfine Interact. 87, 1129
(1994).

\bibitem{MSM} MuonScience: Muons in Physics, Chemistry and Materials, edited by S. L. Lee, S. H. Kilcoyne, and R. Cywinski (Taylor and Francis, Abingdon, 1999).

\bibitem{GML} G. M. Luke, Y. Fudamoto, K. M. Kojima, M. I. Larkin, J.
Merrin, B. Nachumi, Y. J. Uemura, Y. Maeno, Z. Q. Mao, Y.
Mori, H. Nakamura, and M. Sigrist, Nature (London) 394, 558
(1998).
\bibitem{JXY} J. Xia, Y. Maeno, P. T. Beyersdorf, M. M. Fejer, and A.
Kapitulnik, Phys. Rev. Lett. 97, 167002 (2006).

\bibitem{ADH} A. D. Hillier, J. Quintanilla, and R. Cywinski, Phys. Rev. Lett.
102, 117007 (2009).

\bibitem{RPS} R. P. Singh, A. D. Hillier, B. Mazidian, J. Quintanilla, J. F.
Annett, D. M. Paul, G. Balakrishnan, and M. R. Lees, Phys.
Rev. Lett. 112, 107002 (2014).

\bibitem{DSJ} D. Singh, J. A. T. Barker, A. Thamizhavel, D. M. Paul, A. D.
Hillier, and R. P. Singh, Phys. Rev. B 96, 180501(R) (2017).

\bibitem{DSS} D. Singh, Sajilesh K. P., J. A. T. Barker, D. M. Paul,
A. D. Hillier, and R. P. Singh, Phys. Rev. B 97, 100505(R)
(2018).

\bibitem{VKA1} V. K. Anand, A. D. Hillier, D. T. Adroja, A. M. Strydom, H. Michor, K. A. McEwen, and B. D. Rainford Phys. Rev. B 83, 064522 (2011).

\bibitem{LNS} A. J. Drew, S. L. Lee, F. Y. Ogrin, D. Charalambous, N. Bancroft,
D. Mck Paul, T. Takabatake, and C. Baines, Physica B 374-375, 270 (2006).

\bibitem{ZrB} P. K. Biswas, A. D. Hillier, R. P. Singh, N. Parzyk, G. Balakrishnan, M. R. Lees, C. D. Dewhurst, E. Morenzoni, D. McK. Paul, arXiv:1910.09082 (2019).

\bibitem{EMR} E. Morenzoni, R. Khasanov, H. Luetkens, T.Prokscha, A. Suter, N. Garifianov, H. Gluckler, M. Birke, E. Forgan, H. Keller, J. Litterst, Ch. Niedermayer, G. Nieuwenhuys, Physica B 326, 196 (2003).

\bibitem{HLJ} H. Leng, J.-C. Orain,  A. Amato,  Y. K. Huang, and A. de Visser 	arXiv:1910.07248 (2019).


\end{thebibliography}
\end{document}